\documentclass{webofc}
\usepackage[varg]{txfonts}   
\woctitle{?????????}
\begin{document}
\title{CHEOPS: A Transit Photometry Mission for ESA's Small Mission Programme}
%
%

\author{C. Broeg\inst{1,2}\fnsep\thanks{\email{broeg@space.unibe.ch}} \and
        A. Fortier\inst{1,2}\fnsep \and D. Ehrenreich\inst{3} \and
        Y. Alibert\inst{1,2} \and W. Baumjohann\inst{4} \and
        W. Benz\inst{1,2}\and M. Deleuil\inst{5} \and M. Gillon\inst{6} \and A. Ivanov\inst{7}
\and R. Liseau\inst{8}
        \and M. Meyer\inst{9} \and G. Oloffson\inst{10} \and I. Pagano\inst{11} \and G. Piotto\inst{12}
\and D. Pollacco\inst{13} \and D. Queloz\inst{3} 
\and R. Ragazzoni\inst{14}
\and E. Renotte\inst{15} \and M. Steller\inst{4} \and N. Thomas\inst{2}
\and the
        CHEOPS team \inst{16}\fnsep
}

\institute{Center for Space and Habitability, University of Bern
\and Space Research and Planetary Science Division, Physikalisches Institut, Universität Bern, Sidlerstrasse 5, CH-3012 Bern, Switzerland
\and   Observatory of the University of Geneva
\and Space Research Institute, Austrian Academy of Sciences, Graz, Austria
\and Aix Marseille Universit\'e, CNRS, LAM (Laboratoire d’Astrophysique de Marseille) UMR 7326, 13388, Marseille, France
\and Universit\'e de Li\`ege, All\'ee du 6 ao\^ut 17, Sart Tilman, Li\`ege 1, Belgium
\and Swiss Space Center, Ecole Polytechnique F\'ed\'eral de Lausanne, Lausanne, Switzerland
\and Department of Earth and Space Sciences, Chalmers University of Technology, Onsala Space Observatory, 439 92 Onsala, Sweden
\and Institute of Astronomy ETH Z\"urich
\and Department of Astronomy, Stockholm University, 106 91 Stockholm, Sweden
\and INAF Osservatorio Astrofisico di Catania, via S. Sofia 78, 95123 Catania, Italy
\and Dipartimento di Fisica e Astronomia “Galileo Galilei,” Universita` di Padova, v.co dell’Osservatorio 3, I-35122 Padova, Italy
\and Astrophysics Research Centre, School of Mathematics \& Physics, Queen's University, University Road, Belfast BT7 1NN
\and Osservatorio Astronomico di Padova, INAF, vicolo dell’Osservatorio 5, I-35122 Padova, Italy
\and Centre Spatial de Li\`ege
\and http://cheops.unibe.ch
          }

\abstract{%
  Ground based radial velocity (RV) searches continue to discover
  exoplanets below 
  Neptune mass down to Earth mass. Furthermore, ground based transit
  searches now reach milli-mag photometric precision and can discover
  Neptune size planets around bright stars. These searches will
  find exoplanets around bright stars anywhere on the sky,  their
  discoveries representing prime science targets for further study due
  to the proximity and brightness of their host stars.  A mission for
  transit follow-up measurements of these prime targets is currently
  lacking. The first ESA S-class mission CHEOPS (CHaracterizing ExoPlanet
  Satellite) will fill this gap. It will perform ultra-high precision photometric
  monitoring of selected bright target stars almost anywhere on the
  sky with sufficient precision to detect Earth sized transits. It
  will be able to detect transits of RV-planets by photometric
  monitoring if the geometric configuration results in a transit. For
  Hot Neptunes discovered from the ground, CHEOPS will be able to
  improve the transit light curve so that the radius can be determined
  precisely. Because of the host stars' brightness, high precision RV
  measurements will be possible for all targets. All planets observed
  in transit by CHEOPS will be validated and their masses will be
  known. This will provide valuable data for constraining the
  mass-radius relation of exoplanets, especially in the Neptune-mass
  regime. During the planned
  3.5 year mission, about 500 targets will be observed.
  There will be 20\% of open time available for the community to develop new science programmes.
}
\maketitle
\newcommand{\rsun}{\ensuremath{\mathrm R_{\odot}}}
\newcommand{\mearth}{\ensuremath{\mathrm M_{Earth}}}

\section{Introduction}
\label{intro} 
The CHaracterizing ExoPlanet Satellite (CHEOPS) will be the first
mission dedicated to search for transits by means of ultrahigh
precision photometry on bright stars \emph{already known to host planets} in
the super-Earth to Neptune mass range ($1 < M_{planet}/\mearth < 20$). By
being able to point at nearly any location on the sky, it will provide
the unique capability of determining accurate radii for a subset of
those planets for which the mass has already been estimated from
ground-based spectroscopic surveys. The mission will also provide precision radii for new
planets discovered by the next generation ground-based transits
surveys (Neptune-size and smaller).  

While unbiased ground-based searches are well-suited to detect the
transits and fix the ephemerids, CHEOPS is crucial to obtain precise
measurements of planet radii. Knowing {\it where} and {\it when} to
observe makes CHEOPS the most efficient instrument to search for
shallow transits and to determine accurate radii for planets in the
super-Earth to Neptune mass range.

This article focusses on the science objectives and requirements and
briefly outlines the current design.

\section{Science Objectives}
\label{sec:science-objectives}
The main science goal of the CHEOPS mission will be to study the
structure of exoplanets smaller than Saturn orbiting bright
stars. With an accurate knowledge of masses and radii for an
unprecedented sample of planets, CHEOPS will set new constraints on
the structure and hence on the formation and evolution of planets in
this mass range. 
CHEOPS has two main targets: 1) (very) bright stars with a known
planet from RV searches, and 2) bright stars with a known transit
from ground-based transit searches.

\subsection{Mass-radius relation determination}
\label{sec:mass-radius-relation}

The knowledge of the radius of the planet by transit measurements
combined with the determination of its mass through radial velocity
techniques allows the determination of the bulk density of the
planet. Technically, this quantity provides direct insights into the
structure (e.g. presence of a gaseous envelope) and/or composition of
the body (see Fig. 1). Although it is well known that the
determination of planetary structure from bulk density is a highly
degenerate problem, the knowledge of the planet mass and radius
provides enough information to derive a number of basic quantities
relevant to planet structure and hence to formation and evolution to
make them vital measurements for further progress.  

Large ground-based high-precision Doppler spectroscopic surveys
carried out during the last years have identified nearly a hundred
stars hosting planets in the super-Earth to Neptune mass range ($1 < M_{planet}/\mearth < 20$). As search programs continue, the number is
going to increase in the coming years. The
characteristics of these stars (brightness, low activity levels, etc.)
and the knowledge of the planet ephemerids make them ideal targets for
precision photometric measurements from space. The new generation of
ground-based transit surveys (e.g. NGTS), capable of reaching 1 mmag
precision on V < 13 magnitude stars, provide yet another source of
targets. By the end of 2017, NGTS will provide a minimum of 50 targets
in the sub-Saturn size range. 

CHEOPS will determine the mass-radius relation in the planetary mass
range from 20 \mearth{} down to 1 \mearth{} to a precision not achieved
before. In particular, CHEOPS will be able to measure radii to a
precision of 10\% for Neptune-size planets.  

By targeting stars located anywhere on the sky (some biases exist
towards southern hemisphere, where HARPS operates since 2003), which
are bright enough for precise radial velocity follow-up – CHEOPS will
not suffer from the limitations in the planet mass determination
associated with fainter stars. CHEOPS will provide a uniquely large
sample of small planets with well-measured radii, enabling robust bulk
density estimates needed to test theories of planet formation and
evolution. 

\subsection{Identification of planets with atmospheres}
\label{sec:ident-plan-with}

In the core accretion scenario, the core of a planet must reach a
critical mass before it is able to accrete gas in a runaway
fashion. This critical mass depends upon many physical variables,
among the most important of which is the rate of planetesimals
accretion. The determination of the mean planetary density can provide
a lower limit for the mass of the gaseous envelope. For example, a
5-\mearth{} planet composed of 50\% solid terrestrial composition and
50\% water vapour has a radius roughly twice as large as the same-mass
planet with purely terrestrial composition. In light of recent studies
indicating that low density super-Earths with large rocky cores and
hydrogen envelopes may survive outgassing, it seems that the presence of a
significant H/He envelope would have an even more dramatic effect on
the radius due to the reduced molecular weight compared to
water. Similar conclusions can be reached assuming that the planetary
core is composed of pure water ice. Indeed, for a given mass, the
radius of a pure water planet (see Figure 1) represents an upper limit
for the radius of a planet without an envelope. Therefore, a lower
limit to the envelope mass can be derived (as a function of assumed
envelope composition) by matching the observed radius and mass,
assuming a pure water ice core, a composition of the envelope, and a
temperature (corresponding to the equilibrium temperature with the
stellar flux, an adequate assumption if the planet is not located too
close to its star).  

\begin{figure}[htb]
  \centering
  \includegraphics[width=7cm]{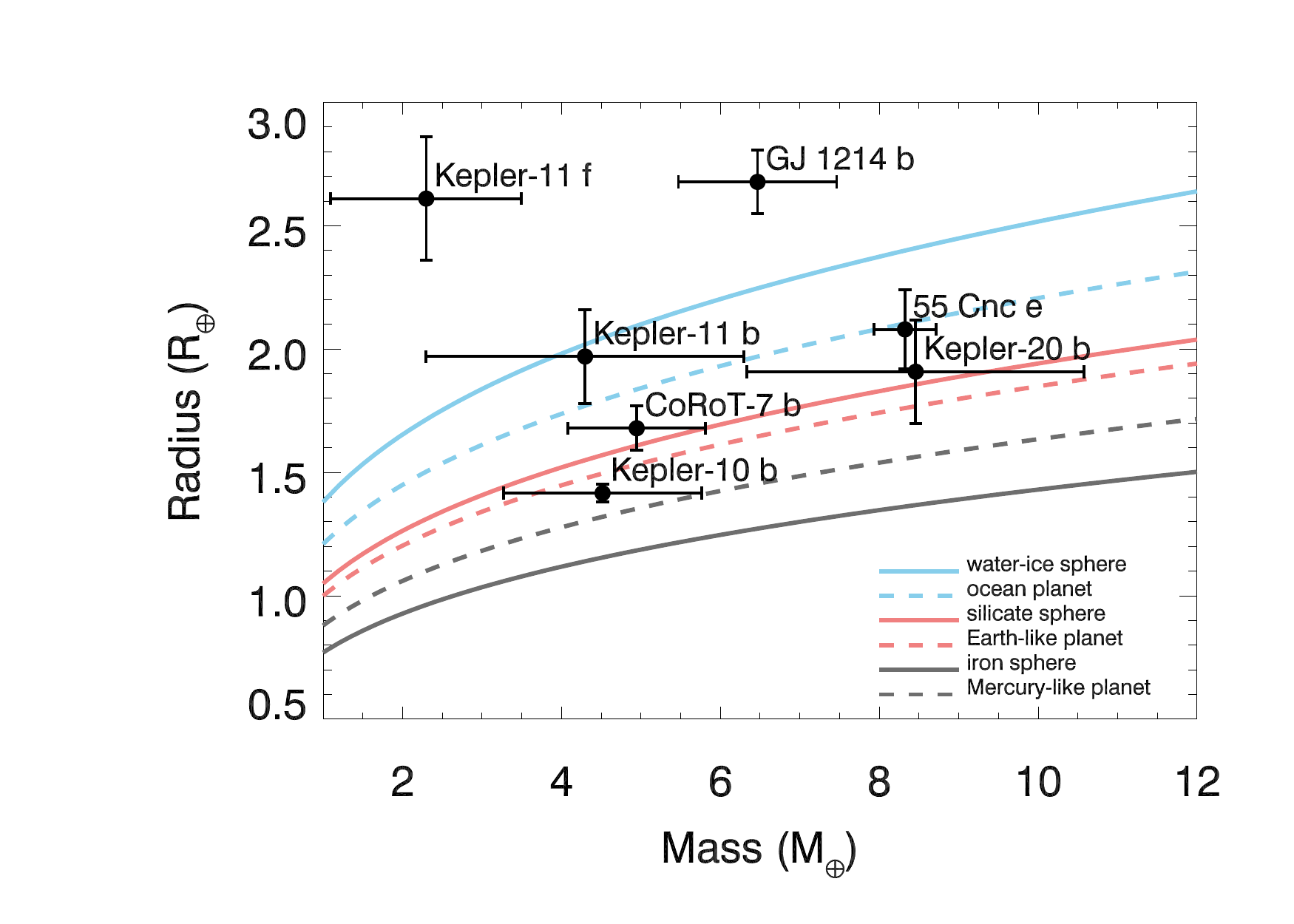}
  \caption{Mass-radius relationship for different bulk composition of
    the planet (Adapted from \cite{wagner}) with superimposed
    known transiting planets where both the mass and the radius of the
    planet have been measured. The size of the boxes indicates the 1-
sigma error on these parameters. So far, in most cases the error bars
are too large to obtain an unambiguous measurement of the bulk
structure of the planets.}
\label{fig1}
\end{figure}

CHEOPS will identify planets with significant atmospheres in a range
of masses, distances from the host star, and stellar parameters. Using
observations of a sample of planets with and without significant
gaseous envelopes, CHEOPS will be able to constrain the critical core
mass (i.e. identify planets that underwent runaway gas accretion, or
those that lost their primordial H-He
atmospheres) as a function of the distance to the star and possibly
stellar parameters (mass, metallicity). This will be especially true
for planets not located extremely close to their stars: if the planet
is too close, evaporation could be an issue thus further complicating
the analysis. 

\subsection{Constraints on planet migration paths}
\label{sec:constr-plan-migr}
It is generally accepted that the envelope masses and compositions of
Uranus and Neptune are directly related to their formation in our own
solar system. Although forming large cores fast enough is a challenge
beyond 10 AU, it is not understood why these two planets did not
succeed in accreting larger amounts of gas. Constraining the gas
fraction for a large sample of Neptune-like planets, but at various
distances to the central star, will shed light on the physical
processes that could produce these types of planetary bodies. Yet even
observations of planets for which it will be impossible to infer
unambiguously the presence of a thick atmosphere (those located below
the blue line in Figure 1) provide strong constraints on formation
models in a statistical sense. There is ample evidence that planets
are not born where they are observed today but that they have migrated
during their formation possibly over large orbital distances. The
present day observed location could therefore have been reached
following different paths depending upon the growth history of the
planet, as well as interactions with the gaseous disc or with other
planets. Each of these paths samples different regions of the
proto-nebula in varying proportions leading to unique combinations
corresponding to the growth history and chemistry appropriate for the
amount of time spent at a given orbital radius. As a result, the bulk
composition, and hence the mean density, will depend upon which track
was followed.  

CHEOPS will provide a sufficiently large sample of planets with
accurate densities to allow discriminating between common groups of
migration paths. In particular, CHEOPS will place constraints on
possible planet migration paths followed during the formation and
evolution of planets where the clear presence of a massive gaseous
envelope cannot be discerned.   

\subsection{Energy transport in hot Jupiter atmospheres}
\label{sec:energy-transport-hot}

The detection of the phase curve provides information on planet
albedos. These have been well measured for CoRoT-1b \cite{snellen} and
HAT-P-7b \cite{borucki2010} with the CoRoT and Kepler 
missions, respectively. The detailed shape and amplitude of the phase
curves represent a powerful tool to study the thermal distribution in
the atmosphere (e.g. HD189733b, \cite{knutson}) and therefore the
physical mechanisms and efficiencies of the energy transport from the
dayside to the night side of the planet. Since this effect can be seen
on any hot Jupiter planet, including non-transiting geometrical
configurations, the number of potential targets amongst hot Jupiters
detected orbiting bright stars is significant. 

CHEOPS will have the capability to detect the phase curve of hot
Jupiters in the optical regime, which will provide information on
planet albedos. CHEOPS will probe the atmospheres of known hot
Jupiters in order to study the physical mechanisms and efficiency of
the energy transport from the dayside to the night side of the planet.  

\subsection{Targets for future spectroscopic facilities}
\label{sec:targ-future-spectr}

Understanding the true nature of super-Earth planets requires not only
precise measurements of their mass and radius, but also a study of
their atmospheric properties. This is only possible for transiting planets
orbiting bright enough stars to permit high signal-to-noise
spectro-photometric observations. This last condition is drastically
more stringent for low-mass planets than for gas giants leading to the
conclusion that only the few dozens of super-Earths that statistically
transit the brightest stars within the solar neighbourhood will ever
be suitable for a thorough characterization with future instruments, e.g. \cite{seager}. This has been nicely demonstrated in the
case of the planet 55 Cnc e. This eight Earth-masses planet is the
only one transiting a star visible to the naked eye. First detected by
Doppler measurements, transits were later detected by the Spitzer and
MOST space telescopes (\cite{demory}, \cite{winn}),
revealing a planet with a size of ~2.1 Earth radii. Owing to the
brightness of its host star (V=6, K=4), very high signal-to-noise
occultation photometry was possible with Spitzer, leading to the
detection of the thermal emission of this super-Earth planet \cite{demory}. 

Earth-like planets are not expected to bear massive atmospheres. Since
the presence of a gaseous envelope (only a few percents in mass) or
icy mantle (above 10\% in mass) has a large effect on the planet
radius and mean density, CHEOPS will be able to discriminate between
telluric, Earth-like planets where life as we know it could blossom,
from other kinds of Earth-mass planets (hydrogen-rich Earths,
ocean-planets), which challenge our understanding of habitability.  

CHEOPS will provide unique targets for future ground- (e.g., E-ELT)
and space-based (e.g., JWST, EChO) facilities with spectroscopic
capabilities. For example, CHEOPS will be able to identify planets
that lack an extended envelope, which are prime targets
for future habitability studies.  

\subsection{Astronomical sources variability studies}
\label{sec:astr-sourc-vari}

CHEOPS will have the capability to provide precise differential
photometric measurements (photometric time series) of a large number
of variable light sources in the Universe. This is regarded as
ancillary science for which observing time will be allocated.  

\section{Science Requirements}
\label{sec:science-requirements}

Here we briefly summarize the key science requirements of the mission.

\subsection{Photometric accuracy}
\label{sec:photometric-accuracy}

Photometric precision for transit detection (RV targets): 
CHEOPS shall be able to detect an Earth-size planet transiting a G5
star (0.9 \rsun) of the 9th magnitude in the V band, with a
signal-to-noise ratio ($S/N_{transit}$) of 10. Since the depth of such a transit is 100
parts-per-million (ppm), this requires achieving a photometric
precision of 10 ppm in 6 hours of integration time. This time
corresponds to the transit duration of a planet with a revolution
period of 50 days. 

Photometric precision for transit characterization (NGTS/ ground based targets): 
CHEOPS shall be able to detect a Neptune-size planet transiting a
K-type dwarf (0.7 \rsun) star of the 12.5th magnitude in the V band
(goal: V=13) with $S/N_{transit}=30$. Such a transit has a
depth of 2500 ppm and last for nearly 3 hours for planets with a
revolution period of 13 days. Hence, a photometric precision of 85 ppm
is to be obtained in 3 hours of integration time. 

\subsection{Sky coverage}
\label{sec:sky-coverage}

Stars with planets detected via Doppler velocimetry: 
50\% of the whole sky should be accessible for 50 days of consecutive observations
 per year and per target with observation
duration longer than 50\% of the spacecraft orbit duration (>50 min
for 100-min spacecraft orbital period). 

Stars with planets detected via ground-based transit surveys: 
25\% of the whole sky, with 2/3 in the southern hemisphere, should be
accessible for 13 days per year and per
target, with observation duration longer than 80\% of the spacecraft
orbit duration (>80 min for 100-min spacecraft orbit). 

\subsection{Temporal resolution}
\label{sec:temporal-resolution-}

Individual exposures should be short enough to avoid saturation on
V$\sim 6.5$ magnitude stars, but the temporal resolution of the measurement should be 1 minute. Full frame images (addition of shorter images when required) will be recorded (and later downloaded) in 1-minute intervals. The time stamp (UTC) uncertainty on the time of exposure should be smaller than 1s.

\subsection{Mission duration}
\label{sec:mission-duration}

Transit detection on bright stars identified by Doppler surveys will
need about 2 days of continuous pointing on target to cope with
uncertainties on radial velocity ephemerids of the longest period
planet (about 3-5\% of the orbital period). With a minimum of 200
targets and 50\% of orbit interruptions, this corresponds to a minimum
total of 800 days of satellite life. 

For NGTS targets a shorter on-target time is required (12 hours). If
one considers 50 targets with a single transit observation and 50
additional targets where 4 transits will be observed and 5 targets
where 10 transits may be required we end up with 150 days. With 20\%
efficiency correction this leads to 180 days of satellite life. 
 
Observations to detect the planets directly in reflected light will be
possible for a handful of hot Jupiters. To obtain a reliable
measurement, disentangled from possible stellar photometric
variability, observations of 3 full planetary orbits are
needed. Assuming a typical 5 days orbital period for hot Jupiter, 15
days of continuous observation are required. Estimating a sample of 5
hot Jupiters for which these observations are required, this
corresponds to 75 days. 

In total these three programs combined require 500 separate target
pointings. Assuming 1 hour per pointing, 10\% margin on each program,
the mission duration is estimated at 1175 days or 3.2 years. Adding to
this duration the open time allocation for carrying out ancillary
science (up to 20\%), the total duration of the CHEOPS mission is
estimated to be 3.5 years.   

\section{Mission Implementation}
\label{sec:mission-summary}

\begin{figure}[htb]
  \centering
  \includegraphics[width=6cm]{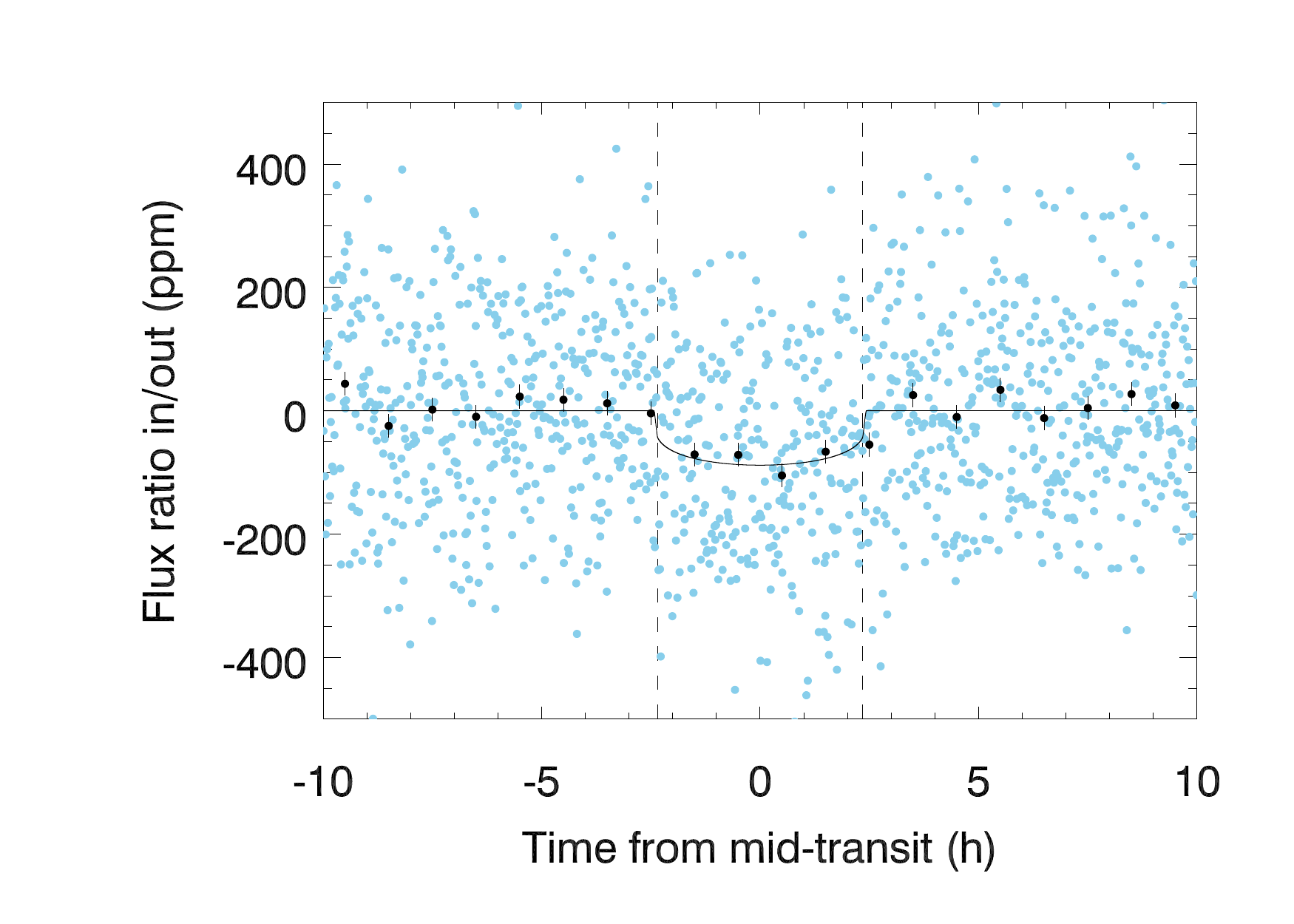}
  \caption{Simulation of a transiting Earth-sized planet with
    a 50 day period orbiting a G5 dwarf star of the 9th magnitude in
    the V band, as observed by CHEOPS. Sampling time is 1 minute and
    photon noise 150 ppm/minute. The black dots indicate 1h-averaged
    photometry. This light curve illustrates a transit detection with
    a $S/N_{transit}=10$.}
\label{fig2}
\end{figure}

To reach its science goals, CHEOPS has to measure photometric signals with a
precision limited by stellar photon noise of 150 ppm/min for a 9th
magnitude star. This corresponds to the transit of an Earth-sized
planet orbiting a star of 0.9 $R_{\odot}$ in 50 days detected with a
S/N$_{transit}$ >10 (100 ppm transit depth). Reaching this ultrahigh
photometric stability on the budget of an S-class mission is
challenging. Figure \ref{fig2} shows the simulated light curve of a Earth-sized
transiting planet.

\subsection{Spacecraft and orbit}
\label{sec:spacecraft-orbit}
The mission will fly a single payload on a small satellite
platform and the total mass of the spacecraft (S/C) will be on the order of 200 kg. To
obtain high photometric stability, thermal stability of the instrument
and straylight suppression from the Earth are design drivers. At the
same time, the observable sky should be maximized. 

To meet the above requirements, the telescope will be orbiting in a
Sun Synchronous Low Earth Orbit (LEO) having a local time of ascending
node (LTAN) of 6 am and 
  an altitude in the range of 620 to 800 km (depending on launch opportunities).
Hence, the satellite
will follow as close as possible the day-night terminator and the
target stars will be above the night side of the Earth. This orbit
also minimizes eclipses and therefore provides a thermally stable
environment. To allow the stringent thermal control of the detector
(see next section), the S/C will be 3-axis stabilized but nadir
locked. Therefore, the payload radiators can always face away from
Earth to cold space. A small sun shield prevents illumination of these
radiators by the Sun, therefore providing a thermally stable
environment for the payload radiators. This orbit allows to fullfill
the science requirements. 

In addition to a thermally stable environment, the instrument requires
high pointing stability: The telescope line-of-sight must remain
stable to 8 arcsec RMS over a 10 hour observing period. This precision
can be achieved on a small platform by including the instrument data
in the attitude control loop. 

\begin{figure}[htb]
  \centering
  \includegraphics[height=4cm]{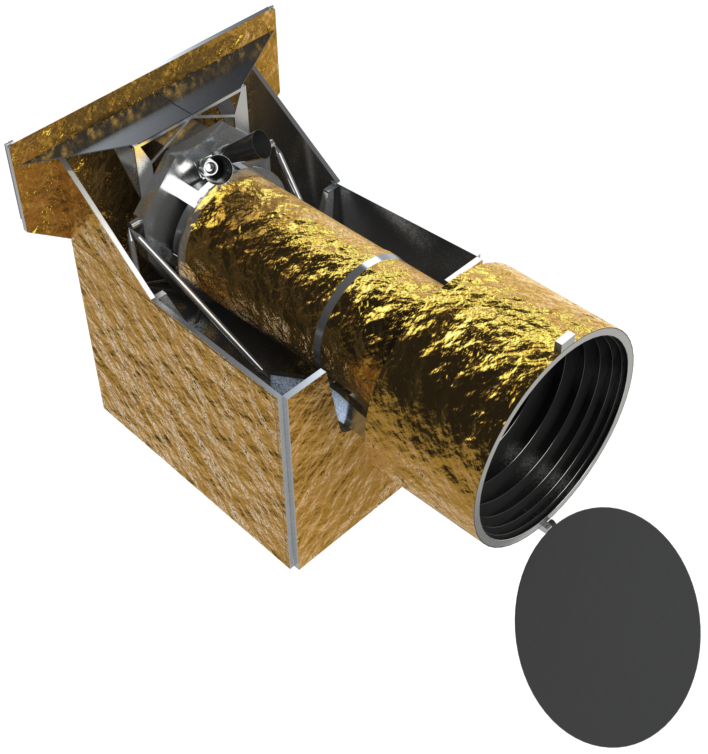}
\hspace{1cm}
\includegraphics[height=4cm]{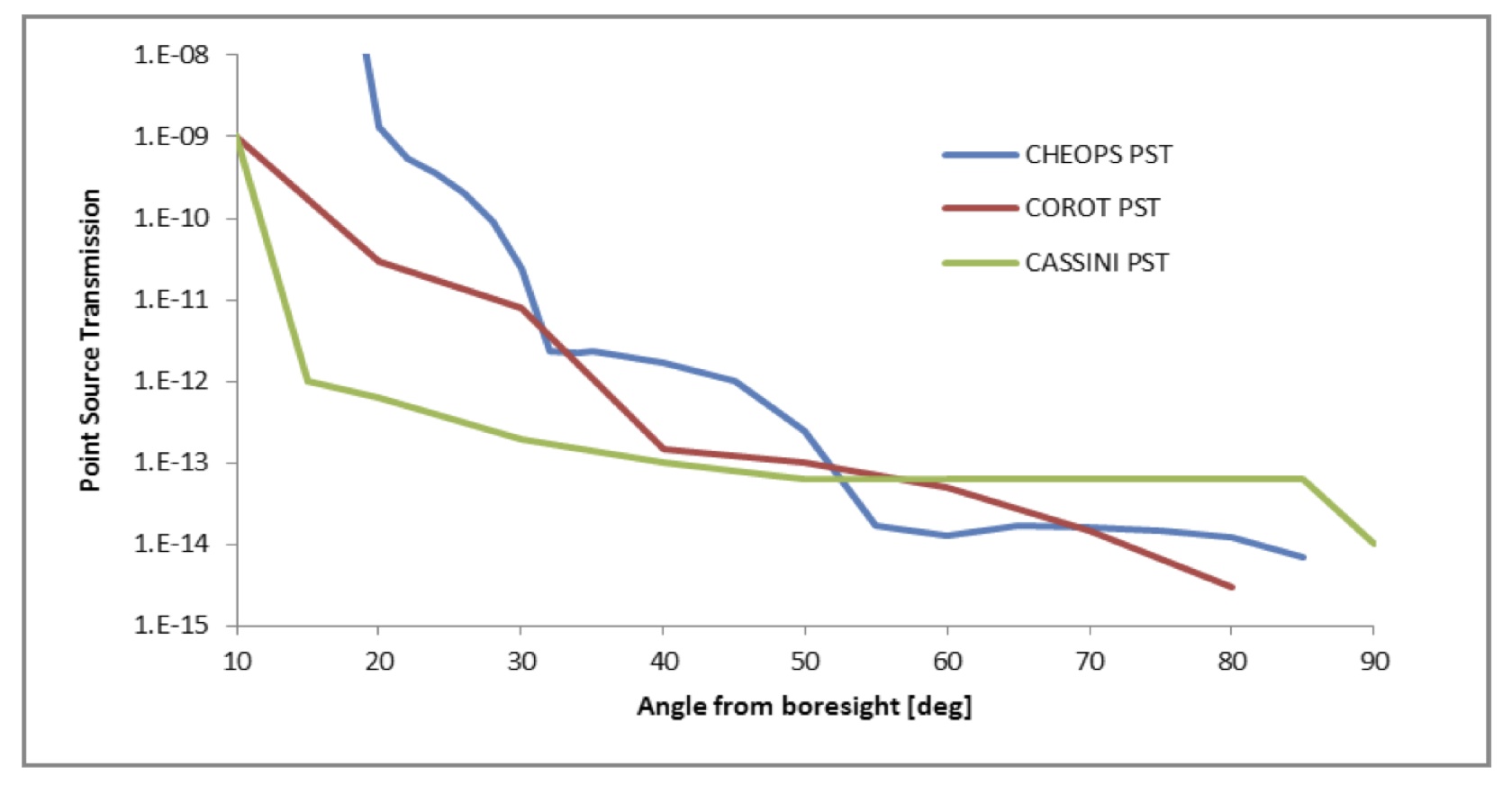}
  \caption{Left: Rendering of the  S/C and payload
    configuration. Right: Stray-light performance of the current
    baffle design in terms of the point source transmission
    function. CHEOPS is optimized for angles larger 35 degrees.}\label{fig3}
\end{figure}

The CHEOPS satellite will observe individual target stars in a track
and stare mode. Following target acquisition of a single star - which
will take less than a minute - the telescope will continuously point
at the target for typically 6-12 hours but up to a few weeks if the
phase modulation of the planet is measured. The telescope operation
will be dominated by many such short pointings, typically only
observing a star when the transit is expected to occur. 
So, from a data point of view, CHEOPS is a simple instrument.  We
baseline an S-band system for 
TM/TC and  data downlink. The S/C
will provide 50 W continuous power for instrument operations and allow
for at least 1Gbit/day downlink. See Figure \ref{fig3} (left) for a
rendering of the S/C.
 
\subsection{Payload}
\label{sec:payload}

The payload is a single instrument: a 30 cm effective aperture
reflecting telescope to observe individual target stars. The major
requirement is photometric stability, 
therefore the detector gain has to be extremely stable and Earth stray
light must be supressed to a very high degree.

The optical design is based on a F/8
Ritchey-Chretien style on-axis telescope and a beam shaper to provide a
de-focussed image of the target star with a point spread function
(PSF) covering an area of ~765 px.
The detector  will be a single frame-transfer back-side illuminated CCD
detector. To achieve the required stability, this detector has to be thermally
stabilized to 5--15 mK at an operating temperature of -40 C. This is achieved by heating against 
cold radiators that are not affected by Solar or Earth radiation.

Being in a LEO, Earth reflected light hast to be prevented from
reaching the detector: a very strong stray light attenuation is
required. 
 An industrial study has led to a
suitable optical design, which  minimizes stray light onto the
detector utilizing a dedicated field stop and a baffling system. This
design meets the requirement of < 1 photon/pixel/second stray light
onto the detector even in the worst case observing geometry on the
baseline orbit (see Figure \ref{fig3}, right, for the achieved PST).

\begin{table}[htb]
  \caption{CHEOPS mission summary}
  \label{tab:msummary}
  \centering
  \begin{tabular}{ll}\hline
    Name&	CHEOPS 
(CHaracterizing ExOPlanet Satellite)\\
Primary science goal&	Measure the radius of planets transiting
bright stars to 10\% accuracy\\
Targets&	Known exoplanet host stars with a V-magnitude < 12.5 anywhere on the sky\\
Instrument&	33 cm reflective on-axis telescope\\
Wavelength&	Visible range : 400 to 1100 nm \\
Detector & 13 $\mu m$ pixel 1k x 1k CCD (baseline: e2v
CCD47-20 AIMO)\\
Total satellite mass & 200 kg\\
Orbit&	LEO sun-synchronous LTAN 6 am (or 6 pm), 620 to 800 km \\
Launch date & 2017\\
Lifetime&	3.5 years\\
Type&	s-class\\\hline
  \end{tabular}
\end{table}

\section{Conclusions}
CHEOPS  will fill the gap in transit follow-up capability for bright stars in the sky. It will target approx. 500 targets of interest in its 3.5 year mission. CHEOPS will:
\begin{itemize}
\item	Determine the mass-radius relation in a planetary mass range for which only a handful of data exist and to a precision never before achieved.

\item	Identify planets with significant atmospheres as a function of their mass, distance to the star, and stellar parameters. \item	Place constraints on possible planet migration paths followed during formation and evolution for planets where the clear presence of a massive gaseous envelope cannot be discerned.

\item	Detect the phase variations of a handful of known Hot Jupiter in order to study the physical mechanisms and efficiency of the energy transport from the dayside to the night side of the planet.

\item	Provide unique targets for future ground- (e.g. E-ELT) and space-based (e.g. JWST, EChO) facilities with spectroscopic capabilities. With well-determined radii and masses, the CHEOPS planets will constitute the best target sample within the solar neighbourhood for such future studies.

\item	Offer up to 20\% of open time to the community to be allocated through competitive scientific review. CHEOPS will have the capability to provide precise photometric measurements (light curves) of a large number of variable light sources in the universe. 
\end{itemize}

%
%
%

\end{document}